\documentclass[epj]{svjour}
\usepackage{graphics}
\begin{document}
\title{Critical Binder cumulant of two--dimensional Ising models}
\author{W. Selke}   
\institute{Institut f\"ur Theoretische Physik, Technische Hochschule
  RWTH Aachen, 52056 Aachen, Germany}
\date{Received: date / Revised version: date}
%
\abstract{
The fourth-order cumulant of the magnetization, the Binder cumulant,
is determined at the phase transition of
Ising models on square and triangular lattices, using Monte
Carlo techniques. Its value at
criticality depends sensitively on
boundary conditions,  details of the
clusters used in calculating the cumulant, and symmetry of the
interactions or, here, lattice structure. Possibilities to
identify generic critical cumulants are discussed.
} 
\maketitle
\section{Introduction}
\label{intro}
In the field of phase transitions and critical phenomena,
the fourth order cumulant of the order parameter \cite{Binder}, the
Binder cumulant $U$, plays an important role. Among others, the
cumulant may be used to compute the critical exponent of the
correlation length, and thence to identify the universality class
of the transition, characterised , e.g., by the values of
the bulk critical exponents \cite{Fisher}.

The value of the Binder cumulant at the transition
temperature in the thermodynamic limit, $U^*$, the critical
Binder cumulant, has received
much attention as well \cite{Privman}, being a measure of 
the deviation of the corresponding distribution function of the order
parameter from a Gaussian function. However, there seem to
be conflicting statements about its 'universality'. For concreteness
and simplicity, let us consider here and in the following results
within the universality class
of the two--dimensional Ising model, with the magnetization
as the order parameter. In particular, in the case of the isotropic
spin-1/2 Ising model with ferromagnetic nearest--neighbour couplings
on a square lattice with $L^2$ spins, the critical cumulant
has been determined very accurately in numerical work, applying
Monte Carlo techniques \cite{Bruce} and transfer-matrix
methods \cite{Bloete} augmented by finite--size extrapolations to the
thermodynamic limit, $L \longrightarrow \infty $. The resulting
value, employing full periodic boundary
conditions, is $U^*= 0.61069...$ \cite{Bloete}. For other
related two-dimensional models on square lattices, including the 
nearest-neighbour XY-model with an easy axis and the spin-1 Ising
model, estimates of $U^*$ have been reported
which seem to be consistent with this value
\cite{Bruce,Bloete,Bin2,Milch,Jan1,Jan2,Schmid,Bin3,Holt}. Actually, the
quoted value for $U^*$ has been sometimes believed to be 'universal',
i.e. to be generic for the two--dimensional Ising universality
class.

On the other hand, the possible dependence of the critical
cumulant, for instance, on boundary conditions has
been noted already by Binder in his pioneering
work \cite{Binder}. Indeed, different values of $U^*$
have been obtained when considering various boundary
conditions, lattice structures (or anisotropic interactions) as
well as aspect ratios, staying in the universality class
of the two--dimensional Ising
model \cite{Privman,Bloete,Burkhardt,Drez,Hilf,Selke}. Some of the
results can be related to each other by suitable transformations. For
instance, applying periodic boundary conditions, the critical
cumulant of the nearest--neighbour Ising model with different
vertical and horizontal couplings may be mapped
onto that of the isotropic model on a rectangular lattice with
aspect ratio $r$ \cite{Bloete,Selke}. Such a scale transformation,
keeping rectangular symmetry and employing periodic boundary
conditions, does not exist, however, for Ising models 
with nearest neighbour and $\it anisotropic$ next--nearest neighbour
interactions on a square lattice (with the triangular lattice being
a special case of that anisotropy \cite{Selke,Chen}). This fact has been
demonstrated by Chen and Dohm \cite{Chen} using renormalization
group arguments, and it
has been confirmed in Monte Carlo
simulations \cite{Selke,Schulte}. It shows
a violation of the two--scale factor universality for finite--size
effects \cite{Priv}, in general, and, specifically, of the
universality of the critical Binder cumulant.    

The aim of this paper is, to study spin-1/2 Ising models with
nearest neighbour interactions on square and triangular 
lattices in order to analyse in a systematic way possible dependences
of the critical Binder cumulant on boundary conditions, clusters
used in calculating the cumulants, and lattice structure (or
anisotropy of the interactions).

The paper is organized as follows: In the next section, the model and
the method are introduced, and the Binder cumulant is
defined. Then, simulational results will be presented, arranged 
according to boundary conditions. Finally, the findings will be
summarized briefly.  

\section{Model and method}
We consider spin-1/2 Ising models on square and triangular
lattices with nearest neighbour ferromagnetic interactions, $J$. The
Hamiltonian reads

\begin{equation}
{\cal H} = - J \sum\limits_{(x,y),(x',y')} S_{x,y} S_{x',y'}
\end{equation}

\noindent
where $S_{x,y}= \pm 1$ is the spin at site $(x,y)$. Sums are taken
over all pairs of nearest--neighour sites $(x,y),(x',y')$. $x$ 
and $y$ refer to symmetry axes of the lattices. Usually, lattices
of linear dimensions $L$ and $K= rL$ will be
simulated, $r$ is the aspect ratio. As indicated above, the
triangular case is isomorphic to an anisotropic Ising
model on a square lattice with nearest--neighbour couplings augmented
by half of the  next--nearest neighbour couplings, also of
strength $J$, along one diagonal direction of
the lattice \cite{Selke,Chen,Berker}.

Our aim is to study the Binder cumulant at the phase transition
temperature $T_c$. For both lattice
structures, the exact critical temperature
is known. For the square lattice, one gets \cite{Onsager} 

\begin{equation}
 k_BT_c/J= 2/\ln(\sqrt{2} +1)= 2.26918...
\end{equation}

\noindent
For the triangular lattice, the critical temperature
is given by \cite{Houtappel}

\begin{equation}
 k_BT_c/J= 2/\ln(\sqrt{3)}= 3.64095..
\end{equation}

The fourth order cumulant of the magnetization, i.e. the Binder
cumulant, for a spin cluster $C$ is defined by \cite{Binder}

\begin{equation}
  U(T,C) = 1- <M^4>_C/(3 <M^2>_C^2)
\end{equation}

\noindent
where $<M^2>_C$ and $<M^4>_C$ denote the second and fourth moments
of the magnetization in that cluster, taking thermal
averages. In principle, clusters of various sizes
or shapes and systems with different boundary conditions may be
studied. In the Ising case, the
cumulant approaches in the thermodynamic
limit (with the cardinal number $|C| \longrightarrow \infty$) the value 2/3 at 
temperatures $T < T_c$, while it tends to zero, reflecting a Gaussian
distribution of the magnetization histogram, at
$T > T_c$ \cite{Binder}. At
$T_c$, $U^*= U(T_c,|C| \longrightarrow \infty $) acquires a nontrivial value,
the critical Binder cumulant.

To study systematically the possible dependence of the critical
cumulant for two-dimensional Ising models on boundary conditions, the choice
of clusters $C$ as well as the lattice type (or, more
basically \cite{Chen}, the anisotropy of interactions), we
performed Monte Carlo simulations for both
lattice types at criticality.

Note that simulational data of high accuracy are needed, to
obtain reliable estimates for $U^*$. We
computed systems of various shapes and sizes, usually
with up to about $4 \times 10^3$ spins. In general, the (moderate) system
sizes already seem to allow for a smooth extrapolation to the
thermodynamic limit. Using the standard Metropolis
algorithm (a cluster flip algorithm becomes significantly
more efficient for larger system sizes), Monte Carlo runs with
up to $10^9$ Monte Carlo steps per site, for the largest systems, were
performed, averaging then over several, up to about ten, of these
runs to obtain final estimates, and to determine the statistical
error bars shown in the figures. We computed not only the
cumulant, but also
other quantities like energy and specific heat, to
check the accuracy of our data. Of course, for sufficiently
small lattices thermal averages may be obtained exactly 
and easily by direct enumeration.

\section{Results}
The critical Binder cumulant depends sensitively on the boundary
conditions. In  fact, Ising systems with periodic and free boundary
conditions will be analysed here. In addition, the Ising model on
a square lattice with mixed, free and periodic, boundary
conditions will be considered.

\subsection{Periodic boundary conditions}
\label{sec:2}
Employing full periodic boundary conditions, with the cluster
$C$ comprising the entire system, $U^*$ has been determined
accurately before, both for square and triangular lattices. For the square
lattice, one gets $U^*_s = 0.61069...$ \cite{Bruce,Bloete,Selke}, and
for the triangular lattice, one finds a slighty different, but
distinct value, $U^*_t = 0.61182...$ \cite{Bloete,Selke}.

Less attention has been paid in the past, however, to different
choices of clusters. In his pioneering work \cite{Binder}, Binder considered
square subblocks for the Ising model on a
square lattice. In particular, for systems of $L^2$ spins, the
clusters then correspond to subblocks of size
$L'^2$, where $L'= bL$, with the subblock factor  $b \le 1$. The
finite--size dependence of the cumulant at criticality has been
discussed as well. For the two--dimensional Ising model, the leading 
correction term to the critical cumulant is
argued \cite{Binder} to behave like $U^*- U_c(T_c, L) \propto 1/L$.

In the original analysis \cite{Binder}, the
subblock sizes $L'$ have been enlargened at fixed $L$. We pursue a
somewhat different strategy in computing cumulants at $T_c$ by
fixing the subblock factor $b$ and then enlargening the linear
dimension of the lattice with $L^2$ spins (applying
periodic boundary conditions). In particular, we set  
$b$= 1, 1/2, 1/4, and 1/8. Some representative
data of our simulations are depicted in Fig. 1. Increasing
the system size $L$, the cumulant at criticality allows for a smooth
and reliable extrapolation to the thermodynamic limit, yielding
$U^*_b$. $U^*_b$ is observed to decrease
with decreasing subblock factor $b$, and we estimate
$U^*_b= 0.5925 \pm 0.0005, 0.577 \pm 0.001$, and $0.568 \pm 0.0015$
for $b$= 1/2, 1.4 , and 1/8, respectively. Plotting now $U^*_b$ against
$b$, we obtain, in the limit $b \longrightarrow 0$, 
the critical cumulant, $U^*_{b=0}= 0.560 \pm 0.002$. This
estimate may be checked by fixing $L'$ and
increasing $L$ to estimate $U^*_{L'}$, see Fig. 2. $U^*_{L'}$ is found
to depend only rather weakly on $L'$, $L' \ge 4$. Taking
into account estimates for $L'$= 4, 8, and 16, we arrive at a
value for $U^*_{b=0}$, which agrees nicely with the one quoted
above. Note that the critical cumulant in the
limit $b \longrightarrow 0$ refers to
arbitrarily large clusters or subblocks being
eventually embedded in their indefinitely larger 'natural' heat bath. In that
sense, the clusters themselves are subject to a 'heat bath boundary
condition'. This boundary condition
is expected to perturb the intrinsic bulk fluctuations
of the magnetization of the cluster only very
mildly. Therefore, $U^*_{b=0}$ may be a candidate for 
a critical Binder cumulant, which is robust against various
modifications of the model. Indeed, within the accuracy
of the simulations, one obtains the same estimate for the critical
cumulant, $b \longrightarrow 0$, when one replaces
the periodic boundary conditions by free boundary conditions, as
will be discussed below.        
                                   
%
\begin{figure}
\resizebox{0.95\columnwidth}{!}{%
  \includegraphics{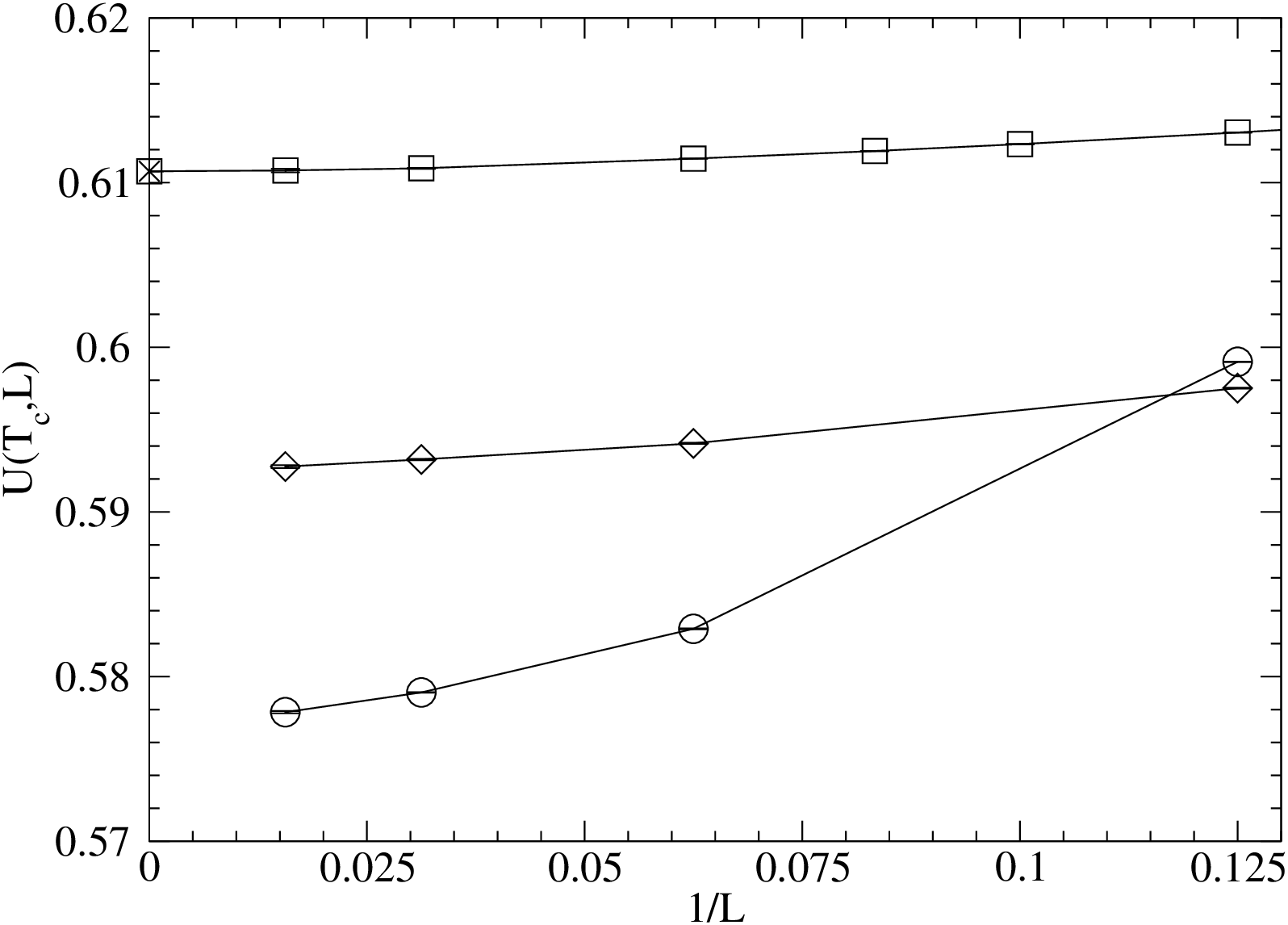}
}
\caption{Binder cumulant $U(T_c,L)$ vs. $1/L$ using subblocks of linear
 size $L'= bL$ for $b$= 1 (squares), 1/2 (diamonds), and 1/4 (circles)
 in the Ising model with
 $L^2$ sites on a square lattice, employing full
 periodic boundary conditions. The square with a cross, at $1/L= 0$,
 refers to the result of Ref. 4.}
\label{fig:1}       
\end{figure}

The critical cumulant for the heat bath boundary
condition of the clusters, $b =0$, is found to depend for rectangular
subblocks, $L' \times r_CL'$, on
their aspect ratio $r_C$, tending to decrease significantly, as
$r_C$ deviates more and more from one.

%
\begin{figure}
\resizebox{0.95\columnwidth}{!}{%
  \includegraphics{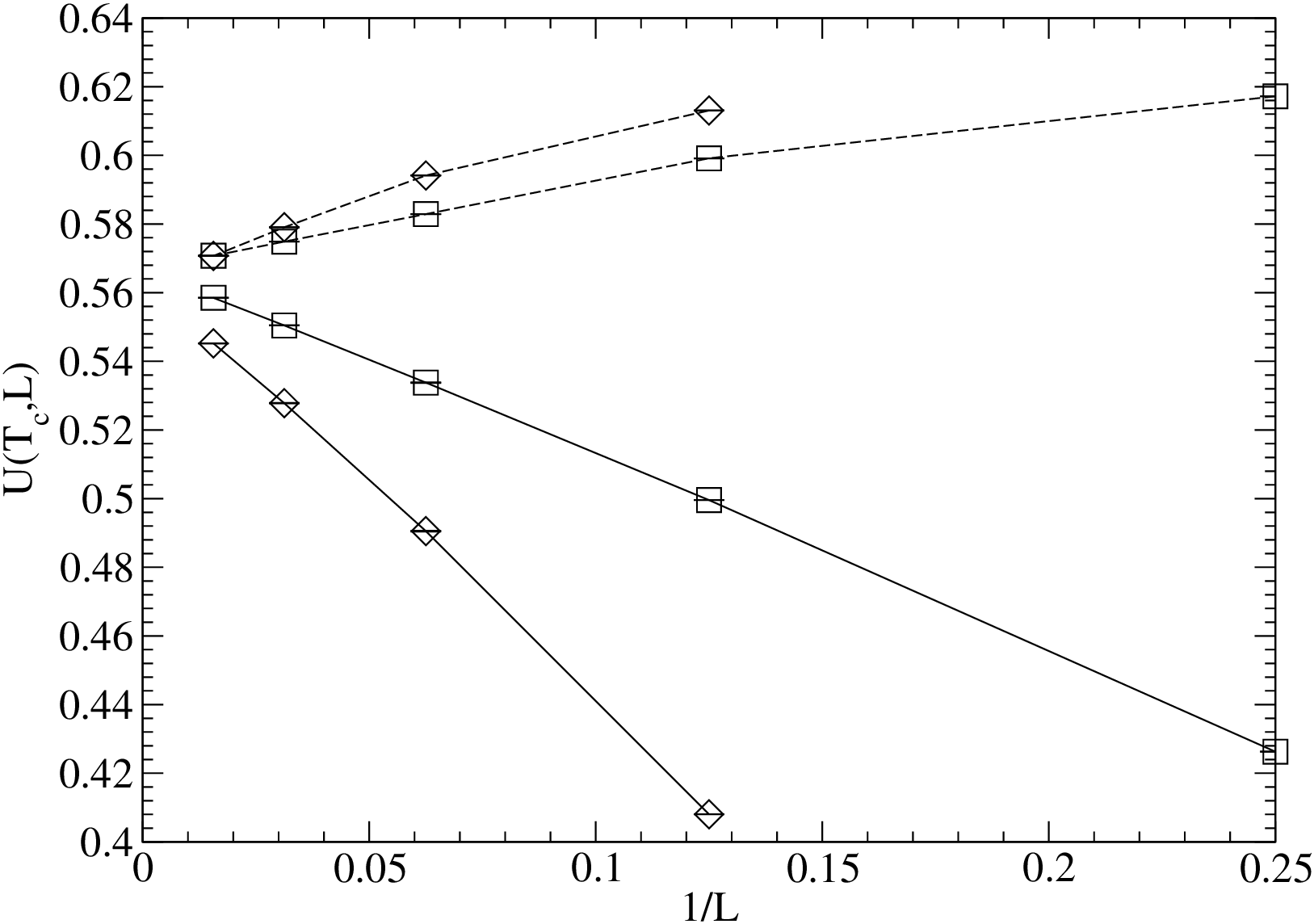}
}
\caption{Fourth--order cumulant $U(T_c,L)$, using subblocks of linear
size $L'$= 4 (squares) and 8 (diamonds) as clusters, versus $1/L$
for square lattices of linear dimension $L$, employing periodic (broken
lines) and free (solid lines) boundary conditions.}
\label{fig:2}       
\end{figure}

Moreover, when going from square to triangular lattices, using periodic
boundary conditions and subblocks of $L'^2$ spins, the critical
cumulant $U^*_{b=0}$ is observed to be very close to that for the square
lattice. The possible difference occurs, perhaps, in the third
digit. But already for $b=1$, the difference
between $U^*_s$ and $U^*_t$ is quite small. Here, the heat bath boundary
condition for the clusters, $b= 0$, tends to reduce such differences
furthermore, and one has to be careful in drawing
definite conclusions. Indeed, on 
physical grounds I tend to believe that also under heat bath boundary
conditions for the subblocks, there is a difference in the value of the
critical cumulant for the triangular and the square lattice, unless
one uses subblocks of special shapes, as will be discussed below.

\subsection{Mixed boundary conditions}
\label{sec:3}

We study the Ising model on a square lattice
consisting of $L$ lines, running from left to right, having
$L$ sites or spins in each line. At the bottom and top, free
boundary conditions are employed, while the left and right hand sides
are connected by periodic boundary conditions. The Hamiltonian, eq. (1), is
slightly extended by still assuming ferromagnetic nearest
neighbour interactions, which now may be different in the
two surface lines at the top and bottom, $J_s$, as compared
to those, $J_b$, in the bulk, i.e. when at least one spin of the 
nearest neighbour pair of spins is not in a
surface line, as usually assumed \cite{Binder4,Diehl,Pleimling,Igloi}. 
                                   
Let us first consider $J_s= J_b$. To compute
the critical cumulant, $U^*_{mixed}$, we take clusters $C$ consisting
of all, $L^2$, thermally excitable spins. Results for various
system sizes are depicted in Fig. 3. Again, the data may be
smoothly extrapolated to the thermodynamic
limit, $L \longrightarrow \infty$, leading to the
estimate $U^*_{mixed}= 0.514 \pm 0.001$.

When varying $J_s/J_b$, the cumulant appears to depend strongly on the
ratio of the surface to the bulk coupling, considering systems
of fairly small sizes, see Fig. 3 for $J_s/J_b$= 0.1. 1.0, and
2.0. However, in the thermodynamic
limit, the critical cumulant seems to approach a unique
value, independent of $J_s/J_b$, as may be inferred also from that
figure.   

It is well known that the critical behaviour of the bulk is 
distinct from that of the
surface \cite{Binder4,Diehl,Pleimling,Igloi}. In
particular, in the two--dimensional case, the vanishing of
the surface magnetization, on approach to $T_c$, is described
by a power law with
an exponent 1/2, while the exponent of the bulk magnetization
is 1/8 \cite{Igloi,Peschel,Selke2}. Thence, it may be interesting
to restrict the clusters $C$ to the surface lines at the bottom
and top of the lattice, in analogy to what has been done before for Ising
films in three dimensions \cite{Landau}. Here, we find that
the critical cumulant tends to vanish in the thermodynamic
limit, reflecting a Gaussian distribution of the histograms for
the surface magnetization. The vanishing may be explained
by the fact that the surface is one--dimensional in our case.

\subsection{Free boundary conditions}
\label{sec:4}

Free boundary conditions allow one to study  arbitrary shapes of
the lattice. Moreover, when compared to periodic
boundary conditions, they are more realistic.

%
\begin{figure}
\resizebox{0.95\columnwidth}{!}{%
  \includegraphics{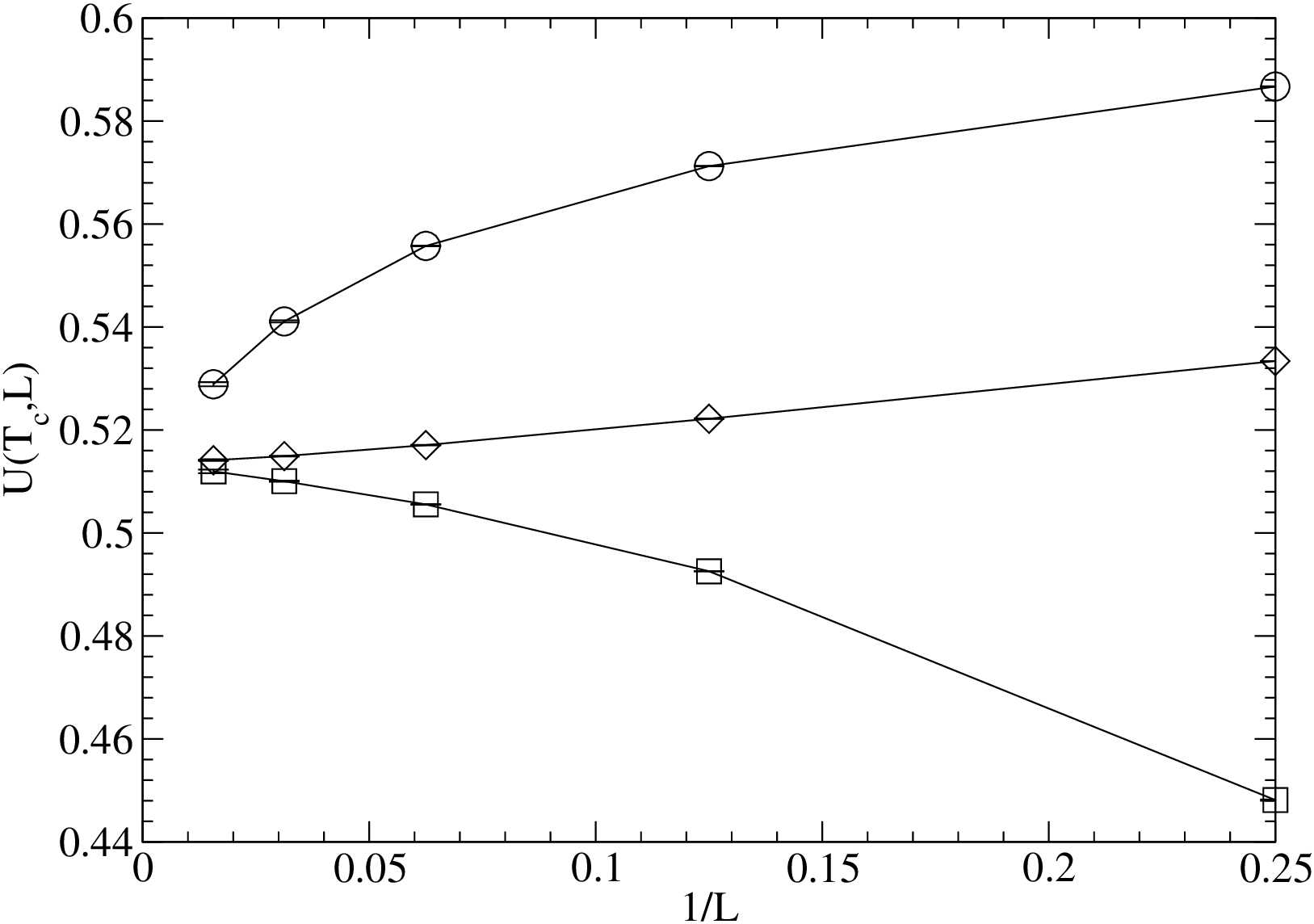}
}
\caption{Binder cumulant $U(T_c,L)$ for the Ising model with
mixed boundary conditions and $L^2$ spins on a square lattice as a
function of $1/L$, varying the
ratio $J_s/J_b$= 0.1 (squares), 1.0 (diamonds), and 2.0 (circles).}
\label{fig:3}       
\end{figure}

Let us first consider square and triangular lattices with $L^2$ spins, i.e.
with aspect ratio $r$= 1, applying free boundary conditions at
the four sides of the system. The clusters $C$ are comprising all
spins, $b$= 1. As may be inferred from Fig. 4, the
critical cumulant, in the thermodynamic limit, may be estimated
from a smooth extrapolation of the simulational data. The resulting
values deviate appreciably for the two different
lattice structures: For the 
square lattice, we find $U^*_{fbc,s}= 0.396 \pm 0.002$, while for
the triangular lattice, we
obtain $U^*_{fbc,t}= 0.379 \pm 0.001$. Accordingly, free boundary
conditions are very useful to show the relevant influence of the
lattice type (or anisotropy of the interactions) on the critical
Binder cumulant. In comparison, the difference in
the critical cumulant for the Ising model on
square and triangular lattices is rather small in the case of
periodic boundary conditions, see above.

The critical cumulant is expected to depend on the aspect ratio, as
we confirmed by considering square lattices with the aspect
ratio $r= 1/2$. The critical cumulant $U^*$ is estimated
to be $0.349 \pm 0.002$.  

We also computed the cumulant, for the
square lattice, with the clusters $C$ being square subblocks of
fixed linear dimension $L'$, to study the effect of
heat bath boundary conditions, with
the subblock factor $b \longrightarrow 0$. As illustrated
in Fig. 2, by increasing $L$, the resulting critical
cumulant $U^*_{fbc}(L')$ tends to approach the same value as in the
case of periodic boundary conditions using the same
subblocks $L'$. The finite--size correction term of the
cumulant has opposite sign for the two boundary
conditions, see Fig. 2. We conclude that there is strong
evidence that the critical cumulant acquires the same value for
free and periodic boundary conditions, when the clusters
are embedded in their natural heat bath. We tend to suggest, that
the very same value holds for other boundary conditions 
as well. On the other hand, as discussed above, the critical cumulant
$U^*$, in the limit $b \longrightarrow 0$, still depends
on the shape of the clusters and, presumably, on the
lattice type (or anisotropy of interaction). 

%
\begin{figure}
\resizebox{0.95\columnwidth}{!}{%
  \includegraphics{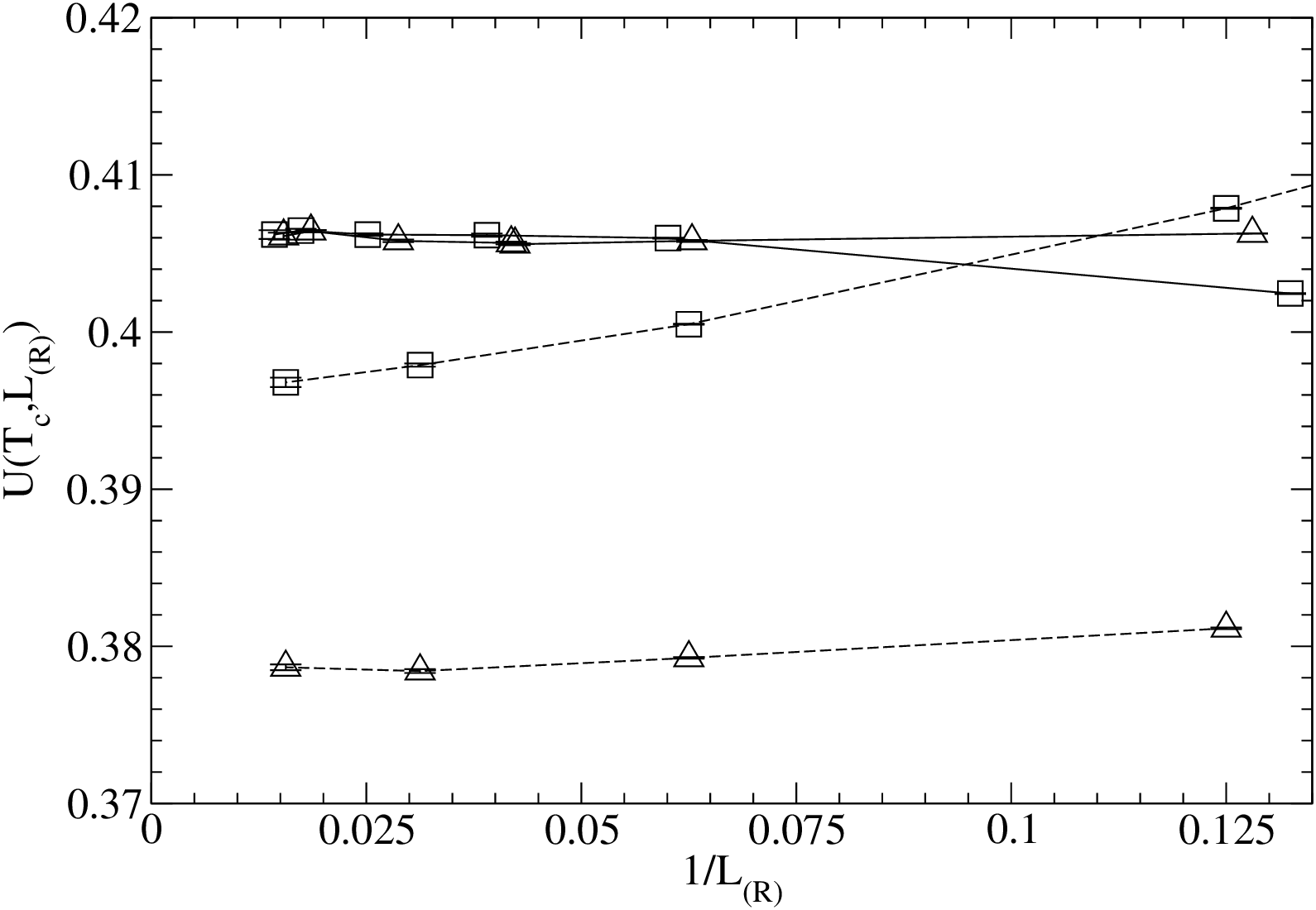}
}
\caption{Binder cumulant $U(T_c,L(_R))$ at criticality
for square (squares) and triangular (triangles) lattices with
free boundary conditions
along two symmetry axes of the lattices (broken lines) and along
discretized circles (solid lines) as a function
of $1/L$  and $1/L_R$, respectively.}
\label{fig:4}       
\end{figure}

The ('non--universal') dependences of $U^*$ may be partly explained
by the fact, that the shape of
the cluster $C$ does not fit to the spatial structure of the
spin correlation function. Indeed, we propose that an
appropriate cluster shape follows from the Wulff
construction at criticality \cite{Abraham}, which determines
equilibrium shapes and preserves the intrinsic symmetry of
the correlations. In the case of square and triangular
lattices, this consideration leads to free boundary conditions
of circular shape. Interpreting
the Ising model on the triangular lattice as a model with anisotropic
next--nearest neighbour interactions on a square lattice, the circle 
would transform into an ellipse, rotated with respect to
the principal axes, on the square lattice. Obviously, for
finite radii, the circular shape may be approximated by a
discretization. More concretely, we define a radius $R$ from
the center of the square or triangular lattice, and we keep all
spins, $N_R$, within this radius as active, thermally excitable
spins, while the remaining spins are set to be equal to
zero. From that construction, an effective linear dimension $L_R$ may
be defined by $L_R= \sqrt{N_R}$ (being proportional to an
effective radius of the cluster). In the thermodynamic
limit, $R \longrightarrow \infty$, one arrives at a perfect
circle. Certainly, an analogous approach is feasible for
clusters with heat bath boundary conditions, $b= 0$, being, however, more
cumbersome, because one had to take, at each given radius, the
thermodynamic limit $L \longrightarrow \infty$.

Simulational data for both lattices with discretized circular
free boundary conditions are depicted in Fig. 4. In contrast to
the case of free boundary conditions along symmetry axes of
the lattices, the critical cumulants for both lattices now
tend to approach close-by, if not identical
values, $U^*_{circle}= 0.406 \pm 0.001$. Of course, further
numerical as well as analytical work will be very useful to
clarify this interesting aspect.

\section{Summary}
\label{sec:1}

In this article, we estimated, using Monte Carlo 
techniques, the critical Binder cumulant $U^*$
for Ising models with nearest neighbour interactions on square and triangular
lattices, employing various boundary conditions, types of
clusters, and aspect ratios. Selected examples are listed in Tab. 1.

\begin{table}
\caption{Selected critical Binder cumulants of the two--dimensional
  Ising model with periodic (pbc), free (fbc), mixed, and heat bath
  (hbbc), $b= 0$, boundary conditions on the square or triangular (tri)
  lattice, considering various system shapes, see text.}
\begin{tabular}{llll}
\hline\noalign{\smallskip}
boundary & lattice & shape & \multicolumn{1}{c}{$U^*$}  \\
\noalign{\smallskip}\hline\noalign{\smallskip}
pbc & square & square & $0.61069...$ \cite{Bloete} \\
pbc & tri & rhombus & $0.61182...$ \cite{Bloete}\\
fbc & square & square & $0.396 \pm 0.002$ \\
fbc & tri & rhombus & $0.379 \pm 0.001$ \\
mixed & square & square & $0.514 \pm 0.001$ \\
hbbc & square & square & $0.560 \pm 0.002$ \\
fbc & square, tri & circle & $0.406 \pm 0.001$ \\
\noalign{\smallskip}\hline
\end{tabular}
\end{table}

In particular, in the case of periodic boundary conditions we
considered square clusters with decreasing subblock
factor $b$. In the limit $b= 0$, we estimate, for the square 
lattice, $U^*_{b=0}= 0.560 \pm 0.002$. The critical
cumulant is observed, when studying clusters of rectangular
shapes, to depend on their aspect ratio. The, presumably, rather
weak dependence of $U^*_{b=0}$ on the lattice structure for these
'heat bath boundary conditions' for the clusters is not resolved 
in our simulations.

For the Ising model with mixed boundary conditions, analysing
square lattices with the aspect
ratio $r= 1$ and clusters comprising all spins, the strength
of the surface coupling is found 
to be irrelevant for the critical cumulant. For
clusters containing only the surface spins, the fluctuations of
the surface magnetization seem to be of Gaussian form with
vanishing $U^*$. This behaviour reflects the fact that the surface is a
one-dimensional object here.

Applying free boundary conditions, the critical Binder cumulants $U^*$
for systems with the aspect ratio $r= 1$ and clusters including all
spins, $b= 1$, are cleary different for square and triangular
lattices ($U^*_{fbc,s}= 0.396 \pm 0.002$,
$U^*_{fbc,t}= 0.379 \pm 0.001$). They
differ significantly from the known corresponding
values for periodic 
boundary conditions. In the limit $b= 0$, we obtain for the
square lattice an estimate for $U^*_{b=0}$ which agrees, within
the error bars, with the one
for periodic boundary conditions. Perhaps most interestingly, employing 
free boundary conditions for clusters of circular form, we
find numerical evidence for a unique value, both for square
and triangular lattices, $U ^*_{circle}= 0.406 \pm 0.001$. In 
general, we suggest that the dependence of the critical
cumulant on the anisotropy of interactions or the lattice
structure may be overcome by
using cluster shapes obtained from the
Wulff construction at criticality.

Certainly, previous standard analyses of the critical
cumulant, using especially periodic boundary conditions with
the subblock factor $b= 1$, are
not invalidated by our study, when they are
interpreted properly. In particular, when comparing critical
cumulants on different models, one has to make sure that
the models satisfy the same symmetries determined by, for
instance, the interactions and/or
lattice structure. In other words, for such analyses universality
of the critical
cumulant holds in a rather restricted sense, when compared
to universality of critical exponents. In any event, care is needed in
applying the critical Binder cumulant when one tries to identify
universality classes or the location of the phase transition.

In general, a finite-size scaling theory including boundary
conditions, system shapes and anisotropy of interactions would be
desirable, extending previous descriptions \cite{Priv,Privman,Antal,Chen}.

Useful discussions with V. Dohm, D. Stauffer, L.N. Shchur, and W. Janke
are gratefully acknowledged.

%
%

\end{document}